\documentclass[12pt]{iopart}
\usepackage{iopams}
\usepackage{graphicx}
\usepackage{float}
\usepackage{subfig}
\usepackage{color,soul}
\begin{document}

\title[Emergence of damped-localized excitations of the Mott state due to disorder]{Emergence of damped-localized excitations of the Mott state due to disorder}

\author{R S Souza$^{1,2,*}$, A Pelster$^{2,\dag}$, and F E A dos Santos$^{1,\ddag}$}

\address{$^1$Departamento de Física, Universidade Federal de São Carlos, 13565-905 São Carlos, SP, Brazil}
\address{$^2$Physics Department and Research Center OPTIMAS, Technische Universität Kaiserslautern, 67663 Kaiserslautern, Germany}
\eads{\mailto{$^*$renan@df.ufscar.br}, \mailto{$^\dag$axel.pelster@physik.uni-kl.de}, and \mailto{$^\ddag$santos@ufscar.br}}
\vspace{10pt}
\begin{indented}
\item[]\today
\end{indented}

\begin{abstract}
A key aspect of ultracold bosonic quantum gases in deep optical lattice potential wells is the realization of the strongly interacting Mott insulating phase. Many characteristics of this phase are well understood, however little is known about the effects of a random external potential on its gapped quasiparticle and quasihole low-energy excitations. In the present study we investigate the effect of disorder upon the excitations of the Mott insulating state at zero temperature described by the Bose-Hubbard model. Using a field-theoretical approach we obtain a resummed expression for the disorder ensemble average of the spectral function. Its analysis shows that disorder leads to an increase of the effective mass of both quasiparticle and quasihole excitations. Furthermore, it yields the emergence of damped states, which exponentially decay during propagation in space and dominate the whole band when disorder becomes comparable to interactions. We argue that such damped-localized states correspond to single-particle excitations of the Bose-glass phase.
\end{abstract}

\section{Introduction}\label{s1}
Ultracold quantum gases in optical lattices offer a unique possibility to investigate strongly interacting many-body systems \cite{jaksch2005cold,bloch2012quantum}. When loaded into a deep lattice potential at commensurate fillings, a cloud of ultracold bosons undergoes a quantum phase transition from a superfluid to a Mott insulating ground state \cite{fisher1989boson,jaksch1998cold,greiner2002quantum}. The latter state is characterized by a gap for quasiparticle and quasihole excitations due to on-site repulsive interactions. In the additional presence of disorder, rare condensate regions emerge inside an insulating background characterizing a Bose-glass ground state \cite{fisher1989boson,damski2003atomic,pollet2009absence,pasienski2010disordered,meldgin2016probing}. Such an effect of spatially random fields can be mimicked by a time-alternating potential where a nonequilibrium granular condensate appears \cite{yukalov2009bose}. The phase transitions between theses states was studied both numerically, with Monte-Carlo simulations \cite{prokof1998worm,astrakharchik2002superfluidity,capogrosso2007phase,gurarie2009phase,soyler2011phase,meier2012quantum,zhang2015equilibrium,ng2015quantum,de2018properties} and stochastic \cite{bissbort2009stochastic,bissbort2010stochastic} as well as local \cite{thomson2016measuring} mean-field techniques, and analytically, where mean-field theory \cite{krutitsky2006mean,buonsante2007mean,pisarski2011application} and field theoretical methods \cite{souza2021green} were applied. Although the nature of the excitations of the Mott state has been the subject of extensive investigation \cite{alon2005zoo,ejima2011dynamic,panas2015numerical,gremaud2016excitation}, the effect of disorder on their energy spectrum remains so far unclear. Furthermore, a concrete characterization of the Bose-glass excitation spectrum is still lacking. Here we demonstrate how the properties of the excitation spectrum in the disordered case can be obtained from the spectral function $A(\bi{k},\omega)$, which generalizes the concept of dispersion relations \cite{abrikosov1963methods,fetter2012quantum,mahan2013many}.

\begin{figure}[t]
\centering
\includegraphics[width=.7\textwidth]{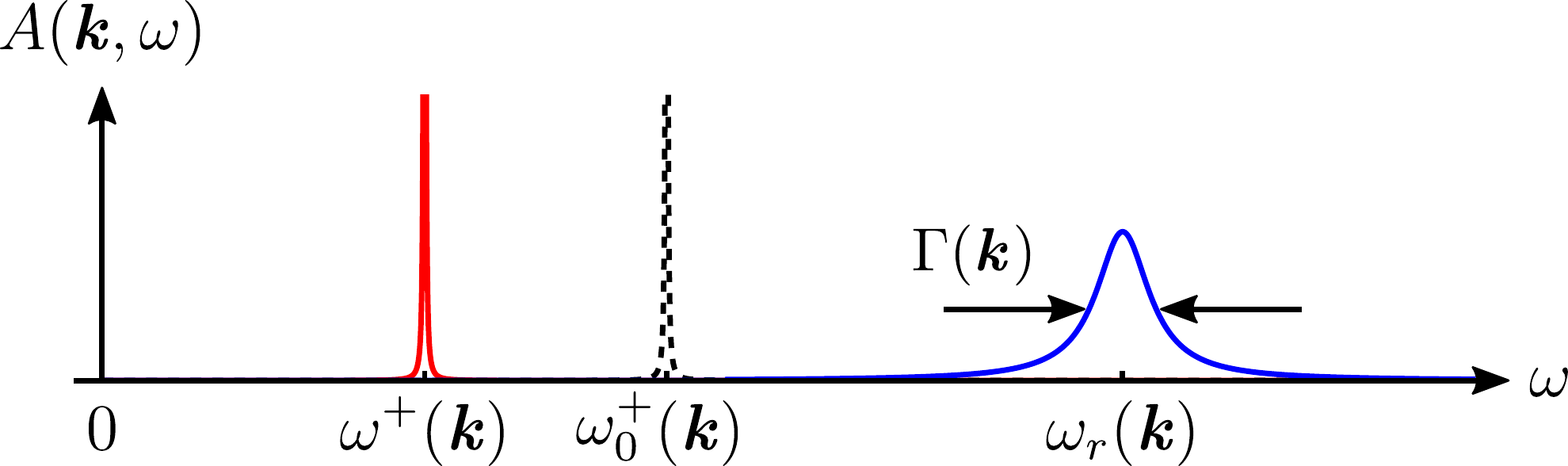}
\caption{\label{fig1} Qualitative sketch of the spectral function $A(\bi{k},\omega)$ for fixed $a|\bi{k}|\ll 1$. The sharp dashed-black peak at $\omega_0^+(\bi{k})$ corresponds to stable excitations of the clean case. Disorder shifts this peak $\omega^+(\bi{k})$ (red) towards lower energies while it also creates damped states with dispersion $\omega_r(\bi{k})$ which appear as a broad peak (blue) of finite width $\Gamma(\bi{k})$.}
\end{figure}

In a perfect lattice, excitations of the Mott state with well-defined wavevector $\bi{k}$ are eigenstates of the underlying Hamiltonian with undamped propagation and finite effective mass inversely proportional to their tunneling energy \cite{mitra2008superfluid,grass2011excitation}. Such states are associated with Dirac distribution peaks of the spectral function centered at the corresponding energy $\hbar\omega^+_0(\bi{k})$, represented by the dashed-black line in Fig.~\ref{fig1}. A detailed analysis of the spectral function in view of characterizing low-energy excitations of the Mott phase can be found, for instance, in Refs.~\cite{panas2015numerical,sengupta2005mott,knap2010spectral,zaleski2012momentum}. When disorder is introduced translational invariance is destroyed. However, if one is not interested in local properties of a single realization of the random potential, one can define a disorder ensemble average which ensures that translational symmetry is recovered \cite{trappe2015semiclassical,signoles2019ultracold}. This motivates the definition of the spectral function as the imaginary part of the averaged single-particle Green's function via \cite{economou2006green}
\begin{equation}\label{eq1}
A(\bi{k},\omega)=-\frac{1}{\pi}\mbox{Im}\langle {\cal G}(\bi{k},\omega+i0^+)\rangle,
\end{equation}
where we consider the analytic continuation from the Matsubara to the real frequency domain consistent with retarded response, i.e. $i\omega_m\rightarrow\omega+i0^+$, and $\langle\cdots\rangle$ stands for the disorder ensemble average. The spectral function \eref{eq1} satisfies both the bosonic positive-definite property and the summation rule \cite{fetter2012quantum}
\begin{equation}\label{eq2}
({\rm sign} \omega)A(\bi{k},\omega)\geq 0,\quad\int^\infty_{-\infty}d\omega A(\bi{k},\omega)=1,
\end{equation} 
also in the presence of disorder.

If disorder and tunneling energy are small compared to the interactions, fluctuations of the particle density become energetically costly and the Mott state still prevails~\cite{fisher1989boson,pollet2009absence,gurarie2009phase}. In this strongly interacting regime, one can apply a perturbative approach based on the strong-coupling expansion. This method was first proposed in the context of lattice bosons by Ref.~\cite{freericks1996strong}, where it was also shown that its first orders provide a remarkable comparison to Monte Carlo simulations. Subsequently, this method served as the basis for the worm algorithm in the Monte Carlo simulations of Refs.~\cite{prokof1998worm,gurarie2009phase,soyler2011phase}, where all orders are taken into account. Recently, the eighth order term was calculated in Ref.~\cite{wang2018high}, demonstrating that the strong-coupling approximation works extremely well in predicting the quantum phase boundary. In this work, we use a strong-coupling approach based on field-theoretical methods, such as those developed in Refs.~\cite{dos2009quantum,bradlyn2009effective,souza2021green}, to construct an expansion of the Green's function for small tunneling energy values, considering only tree-level corrections. This approach leads to a mean-field phase boundary equivalent to the one found in~\cite{krutitsky2006mean,souza2021green}. Therefore, for sufficiently small tunneling energy, such a perturbative method is sufficient to describe the low-energy excitations in the disordered case.

In what follows we report that, similarly to the clean case, stable excitations are still present and are associated with sharp peaks in the spectral function. However, as the region occupied by the Mott states in the phase diagram shrinks due to increasing disorder \cite{krutitsky2006mean,souza2021green}, the gap for these stable states decreases leading to a shift of its dispersion $\omega^+(\bi{k})$ towards lower energies. Additionally, we find that a broad peak of finite width $\Gamma(\bi{k})$ emerges in the spectral function due to scattering effects with the random potential. Such a broad distribution corresponds to damped states with dispersion $\omega_r(\bi{k})$. The qualitative features of our results for the quasiparticle branch of the spectrum are schematically depicted in Fig.~\ref{fig1}. For the quasihole branch of the spectrum, which occurs at negative energies, the qualitative results are analogous. For strong disorder, the damped states occupy the whole spectrum. In this limit, the dispersive nature of the excitations ceases to exist implying a large distribution in momentum, which is typical of localized states. We argue that the new set of damped-localized states correspond to single-particle excitations of the Bose glass. Moreover, we demonstrate that in the case of bounded uniform randomness the effective mass of stable excitations increases with increasing disorder. At the same time, the lifetime of the damped states also increases.

We proceed by first deriving a hopping expansion to the single-particle Green's function in Sec. 2. In Sec. 3 we discuss the characteristics of the spectral function. We then consider in Sec. 4 the case of a uniform disorder distribution and obtain the band structure for the case where the average particle density at each site is $n=0$ and $n=1$. In Sec. 5 we discus the spatio-temporal profile of the Green's function. Finally, we summarize the implications of our results and conclude in Sec. 6.  

\section{Hopping expansion}\label{s2}

We start our analysis by defining the Bose-Hubbard Hamiltonian, which describes spinless bosons with short-range interactions on a lattice potential 
\begin{equation}
\label{eq3}
\hat{H}=\sum_i\Bigg[\frac{U}{2}\hat{n}_i(\hat{n}_i-1)+(\epsilon_i-\mu)\hat{n}_i\Bigg]-\sum_{ij}J_{ij}\hat{a}^\dagger_i\hat{a}_j.
\end{equation}
Here $\hat{a}^\dagger_i$ and $\hat{a}_i$ satisfy standard bosonic commutation relations, $\hat{n}_i=\hat{a}^\dagger_i\hat{a}_i$ is the particle number operator, $\mu$ denotes the chemical potential. In addition, $U$ represents the on-site interaction and $J_{ij}$ stands for the coupling between neighboring sites, i.e., it takes the value $J$ only when $i$ and $j$ are first neighbors and vanishes otherwise. Furthermore, we assume that the local imperfections $\epsilon_i$ are uncorrelated at different sites and randomly distributed over the lattice according to a distribution $p(\epsilon_i)$ bounded in the interval $[-\Delta/2,\Delta/2]$. Therefore, $\Delta$ represents the scale associated with the energy shift caused by the random potential on the lattice.

In the decoupled limit of $J=0$, the Hamiltonian $\hat{H}_0=\sum_i[\frac{U}{2}\hat{n}_i(\hat{n}_i-1)+(\epsilon_i-\mu)\hat{n}_i]$ is diagonal in the particle number operator basis. Considering the hopping term $\hat{V}=-\sum_{ij}J_{ij}\hat{a}^\dagger_i\hat{a}_j$ as a perturbation, we use the Dirac interaction picture representation to write the imaginary-time evolution operator as 
\begin{equation}\label{eq4}
{\rm e}^{-\tau\hat{H}}{\rm e}^{\tau^\prime\hat{H}}={\rm e}^{-\tau\hat{H}_0}\hat{\cal U}(\tau,\tau^\prime){\rm e}^{\tau^\prime\hat{H}_0},
\end{equation}
\begin{equation}\label{eq5}
\hat{\cal U}(\tau,\tau^\prime)=\hat{\cal T}{\rm exp}\Bigg[-\int^\tau_{\tau^\prime}d\tau_1\hat{V}(\tau_1)\Bigg],
\end{equation}
where we have set $\hbar=1$. We employ the standard definition for the imaginary-time dependent operator $\hat{V}(\tau)={\rm e}^{\tau\hat{H}_0}\hat{V}{\rm e}^{-\tau\hat{H}_0}$ and $\hat{\cal{T}}$ is the time-ordering operator. Thus, we define the single-particle Green's function as
\begin{equation}\label{eq6}
{\cal G}_{ij}(\tau)=-\frac{\mbox{Tr}\{{\rm e}^{-\beta\hat{H}_0}\hat{\cal T}[\hat{U}(\beta,0)\hat{a}_i(\tau)\hat{a}_j^\dagger(0)]\}}{\mbox{Tr}[{\rm e}^{-\beta\hat{H}_0}\hat{U}(\beta,0)]}.
\end{equation}
One can obtain a perturbative approximation to the above quantity by expanding the exponential of the operator $\hat{\cal U}$ in powers of the hopping $\hat{V}$ and evaluating the corresponding traces with respect to the equilibrium ensemble of the unperturbed Hamiltonian $\hat{H}_0$ for fixed $\epsilon_i$. This approximation is analogous to the strong-coupling expansion used in Ref. \cite{freericks2009strong} considering additionally frozen disorder. A similar method was applied in Ref. \cite{freericks1996strong} which provided a significant comparison against Monte-Carlo calculations for the prediction of the Mott-lobes phase boundary both in the clean case and in the presence of box disorder.

Each term in the expansion of \eref{eq6} can be associated with a diagram corresponding to the path of an excitation which starts from site $j$ at imaginary time zero and hops along the lattice reaching site $i$ at imaginary time $\tau$ \cite{dos2009quantum,bradlyn2009effective,souza2021green}. After combining the expansions of the numerator and the denominator of \eref{eq6} order by order, one can write the remaining terms only using the connected parts of the Green's function. This result follows from the so-called linked cluster theorem \cite{mahan2013many,freericks2009strong,fradkin2021quantum}. Considering only the first-order correction in the hopping expansion we get
\begin{equation}\label{eq7}
{\cal G}_{ij}(\tau)=\delta_{ij}g_i(\tau)+J_{ij}\int^\beta_0d\tau_1 g_i(\tau-\tau_1)g_j(\tau_1)+\cdots,
\end{equation}
where we define the unperturbed Green's function as 
\begin{equation}\label{eq8}
g_i(\tau)=-\frac{\mbox{Tr}\{{\rm e}^{-\beta\hat{H}_0}\hat{\cal T}[\hat{a}_i(\tau)\hat{a}_j^\dagger(0)]\}}{\mbox{Tr}[{\rm e}^{-\beta\hat{H}_0}]}.
\end{equation}
By transforming into the Matsubara frequency domain  
\begin{equation}\label{eq9}
{\cal G}_{ij}(i\omega_m)=\int^\beta_0d\tau {\cal G}_{ij}(\tau){\rm e}^{i\omega_m\tau},
\end{equation}
we get that \eref{eq7} can be rewritten as
\begin{equation}\label{eq10}
{\cal G}_{ij}(i\omega_m)=\delta_{ij}g_i(i\omega_m)+J_{ij}g_i(i\omega_m)g_j(i\omega_m)+\cdots.
\end{equation}
In the zero-temperature limit, the Matsubara frequencies become continuous and the unperturbed Green's function reads
\begin{equation}\label{eq11}
g_i(i\omega)=\frac{n+1}{i\omega+\mu-\epsilon_i-Un}-\frac{n}{i\omega+\mu-\epsilon_i-U(n-1)},
\end{equation}
where $n\in\mathbb{N}_0$ is the average particle density that minimizes the energy at each site. Note that the unperturbed Green's function is characterized by two simple poles that correspond to stable quasiparticle and quasihole states. 

The presence of disorder leads to randomly distributed shifts in the lattice potential. Such shifts act as scattering centers for the excitations. By averaging over all possible disorder configurations and considering independent scattering events for the propagation of the single-particle excitations, we obtain that the inverse of the Green's function can be written as
\begin{equation}\label{eq12}
\langle {\cal G}_{ij}(i\omega)\rangle^{-1}=\delta_{ij}\frac{1}{\langle g_i(i\omega)\rangle}-J_{ij}+\cdots,
\end{equation}
where the disorder average is defined according to $\langle \cdots\rangle=\prod_i\int^{\infty}_{-\infty} d \epsilon_i \cdots p(\epsilon_i)$. By using the Fourier transform 
\begin{equation}\label{eq13}
\langle {\cal G}(\bi{k},\bi{k}^\prime;i\omega)\rangle^{-1}=\sum_{ij}\langle G_{ij}(i\omega)\rangle^{-1}\e^{-i(\bi{k}\cdot\bi{r}_i-\bi{k}^\prime\cdot\bi{r}_j)},
\end{equation}
we invert \eref{eq12} exactly which gives
\begin{equation}\label{eq14}
\langle {\cal G}(\bi{k},\bi{k}^\prime;i\omega)\rangle=\Bigg(\frac{2\pi}{a}\Bigg)^D\frac{\delta(\bi{k}-\bi{k}^\prime)}{\langle g_i(i\omega)\rangle^{-1}-J(\bi{k})}.
\end{equation}
Note that this result has the form of a Dyson equation, where the self-energy is given by the dispersion $J(\bi{k})=2J\sum_{\alpha=1}^D \cos(a k_\alpha)$ for a $D$-dimensional hypercubic lattice with spacing $a$ between sites. Such a perturbative result is equivalent to the tree-level approximation used, for instance, in Ref.~\cite{souza2021green}. Thus, it amounts to considering an infinite amount of paths for an excitation created at a given site $j$ with wavevector $\bi{k}$ to hop to a different site $i$ and be annihilated with the same wavevector $\bi{k}$ by suffering independent scattering processes at each site. It should be noted that the partial summation result of \eref{eq14} tends to underestimate fluctuations in the hopping expansion. Therefore, it is considered to be accurate only for sufficiently small tunneling energy, where second-order loop corrections can be disregarded. In the absence of disorder, it has been demonstrated that this result can effectively describe the collective excitations of superfluid ground states. Specifically, it predicts a gapless Goldstone mode and a gapped amplitude mode. This was achieved in Ref. \cite{bradlyn2009effective} through an effective action approach. Here, however, we concentrate on the analysis on the strong interacting limit where the ground state is a Mott insulator.

We therefore define the retarded Green's function by taking the analytic continuation from the Matsubara to real frequency domain, $i\omega\rightarrow\omega+i 0^+$,
\begin{equation}\label{eq15}
\langle {\cal G}(\bi{k},\omega+i 0^+)\rangle=\Big[\langle g_i(\omega+i 0^+)\rangle^{-1}-J(\bi{k})\Big]^{-1},
\end{equation}
where the disorder average of the unperturbed Green's function at zero temperature is given by
\begin{equation}\label{eq16}
\fl
\eqalign{\langle g_i(\omega+i 0^+) \rangle = &\mathcal{P}\int^{\infty}_{-\infty}  d \epsilon_i\Bigg[\frac{(\epsilon_i-\omega-\mu-U)p(\epsilon_i)}{(\omega+\mu-Un-\epsilon_i)(\omega+\mu-U(n-1)-\epsilon_i)}\Bigg]
\\
&+i\pi\bigg[np(\omega+\mu-U(n-1))
-(n+1)p(\omega+\mu-Un)\bigg]},
\end{equation}
with ${\cal P}$ denoting the principal value of the integral of the first term.
Note that the imaginary part of \eref{eq16} is controlled by the disorder distribution. The clean case is recovered by setting $p(\epsilon_i)=\delta(\epsilon_i)$. Hence, we observe that the disorder averaging amounts to summing up infinitely many simple poles, each one coming from a single realization of the unperturbed Green's function, thus transforming the simple-pole structure of the excitations of the clean case into a branch cut. In the strong disorder limit $\Delta>U$, the particle density $n$ never sticks to an integer value implying that no Mott insulating state exists \cite{fisher1989boson,pollet2009absence}. Thus, we focus on the case of a Mott state where $\Delta$ is finite and smaller than the on-site interaction strength, i.e., $\Delta<U$. In the next section we analyze the implications of our results \eref{eq15} and \eref{eq16} on the spectral function. 

\section{Spectral function}

We obtain the spectral function by applying the definition \eref{eq1} to \eref{eq15}
\begin{equation}\label{eq17}
A(\bi{k},\omega)=\frac{-\frac{1}{\pi}\mbox{Im}\langle g_i(\omega)\rangle}{[1-J(\bi{k})\mbox{Re}\langle g_i(\omega)\rangle]^2+[J(\bi{k})\mbox{Im}\langle g_i(\omega)\rangle]^2}.
\end{equation}
The states of the excitations are associated with peaks of the spectral function. Assuming that $\mbox{Im}\langle g_i(\omega)\rangle$ is a smooth function of $\omega$, the peaks of $A(\bi{k},\omega)$ coincide with the frequencies that satisfy
\begin{equation}\label{eq18}
1-J(\bi{k})\mbox{Re}\langle g_i(\omega)\rangle=0.
\end{equation}
Thus, the solutions of \eref{eq18} for fixed $n$ correspond to the dispersion relations of the respective excitations inside the Mott state. 

If $\mbox{Im}\langle g_i(\omega)\rangle$ approaches zero, the spectral function will only be non-zero if \eref{eq18} is satisfied. When this condition is met, the spectral function can be expressed as
\begin{equation}\label{eq19}
A^\pm(\bi{k},\omega)=\pm\frac{\delta(\omega-\omega^\pm(\bi{k}))}{|J(\bi{k})^2\partial_\omega\mbox{Re}\langle g_i(\omega^\pm(\bi{k}))\rangle|},
\end{equation}
where the plus (minus) sign relates to stable quasiparticle (quasihole) excitations. The vanishing width of the $\delta$-distribution characterizes a stable propagation of these excitations. From the dispersion relations $\omega^\pm(\bi{k})$ we define the effective-mass tensor with components \cite{kittel1996introduction}
\begin{equation}\label{eq20}
(m^*)_{ij}^{-1}=\Bigg(\frac{\partial^2\omega^\pm(\bi{k})}{\partial k_i\partial k_j}\Bigg)\Big\vert_{\textbf{\textit{k}}=\textbf{\textit{0}}}.
\end{equation}

On the other hand, when $\mbox{Im}\langle g_i(\omega)\rangle$ is finite, resonances occur whenever \eref{eq18} is fulfilled. Near such resonances $\omega\approx\omega_r(\bi{k})$ we expand \eref{eq17} according to
\begin{equation}\label{eq21}
\frac{\mbox{Im}\langle g_i(\omega)\rangle}{|1-J(\bi{k})\langle g_i(\omega)\rangle|^2}\approx\frac{\Omega(\bi{k})\Gamma(\bi{k})}{[\omega-\omega_r(\bi{k})]^2+\Gamma(\bi{k})^2}.
\end{equation}
This consists of approximating the spectral function by a Cauchy-Lorentz distribution in that region. Such an expansion yields an estimate for the renormalization factor 
\begin{equation}\label{eq22}
\Omega(\bi{k})=|J(\bi{k})^2\partial_\omega\mbox{Re}\langle g_i(\omega_r(\bi{k}))\rangle|^{-1},
\end{equation}
and for the width of the distribution
\begin{equation}\label{eq23}
\Gamma(\bi{k})=\Big\vert\frac{\mbox{Im}\langle g_i(\omega_r(\bi{k}))\rangle}{\partial_\omega\mbox{Re}\langle g_i(\omega_r(\bi{k}))\rangle}\Big\vert,
\end{equation}
whose inverse
\begin{equation}\label{eq24}
\tilde{\tau}(\bi{k})=\frac{1}{\Gamma(\bi{k})},
\end{equation}
is interpreted as the lifetime of these states \cite{economou2006green}. As the region where $\mbox{Im}\langle g_i(\omega)\rangle$ is finite is controlled by the disorder distribution $p(\epsilon_i)$, the emergence of these damped states is directly linked to the presence of the disorder potential. Notice that these resonances appear as poles in the lower half plane of the retarded Green's function $\langle {\cal G}(\bi{k},\omega+i0^+)\rangle$. 

Therefore, for each Mott lobe with fixed $n$, the energy band splits into two regions. The first one is associated with stable states. The second one emerges due to the disorder. It corresponds to damped states with a characteristic lifetime $\tilde{\tau}(\bi{k})$. Thus, we separate the corresponding terms of the spectral function as 
\begin{equation}\label{eq25}
\fl
\eqalign{
A(\bi{k},\omega)=&\frac{\delta(\omega-\omega^+(\bi{k}))}{|J(\bi{k})^2\partial_\omega\mbox{Re}\langle g_i(\omega^+(\bi{k}))\rangle|}
-\frac{\delta(\omega-\omega^-(\bi{k}))}{|J(\bi{k})^2\partial_\omega\mbox{Re}\langle g_i(\omega^-(\bi{k}))\rangle|}
\\
&+\frac{(n+1)p(\omega+\mu-Un)}{[1-J(\bi{k})\mbox{Re}\langle g_i(\omega)\rangle]^2+\pi^2(n+1)^2J(\bi{k})^2p(\omega+\mu-Un)^2}
\\
&-\frac{np(\omega+\mu-U(n-1))}{[1-J(\bi{k})\mbox{Re}\langle g_i(\omega)\rangle]^2+\pi^2n^2J(\bi{k})^2p(\omega+\mu-U(n-1))^2}.
}
\end{equation}
Note that we have made no assumption so far on the specific value of $n$ nor on the concrete shape of the bounded distribution $p(\epsilon_i)$. Therefore, all the conclusions drawn so far can be considered to be generally valid within each Mott lobe, subject to the limitations of our approximations. We remark that, even though our method is better suited for high number of dimensions, it is still applicable in $D=1$ for sufficiently small $J$ \cite{dos2009quantum}.

We now turn our attention to the case of a uniform distribution.

\section{Uniform disorder distribution}

To better demonstrate the above described results, we investigate further the case of uniform disorder distribution $p(\epsilon_i)=\frac{1}{\Delta}[\Theta (\epsilon_i+\frac{\Delta}{2})-\Theta (\epsilon_i-\frac{\Delta}{2})]$, where $\Theta(x)$ is the Heaviside function. Such a distribution can be generated, for instance,  by customizing the intensity in speckle laser experiments \cite{bender2018customizing}.  In this case, \eref{eq18} reduces to
\begin{equation}\label{eq26}
\eqalign{\Big\vert\omega&+\mu-Un-\frac{\Delta}{2}\Big\vert^{n+1}\Big\vert\omega+\mu-U(n-1)+\frac{\Delta}{2}\Big\vert^n
\\
&-\mathrm{e}^{\Delta/J(\bi{k})}\Big\vert\omega+\mu-Un+\frac{\Delta}{2}\Big\vert^{n+1}\Big\vert\omega+\mu-U(n-1)-\frac{\Delta}{2}\Big\vert^n
=0.}
\end{equation}
Note that not all solutions of \eref{eq26} are solutions of \eref{eq18}. For $n\geq 1$ we expect from the above arguments that only four solutions of \eref{eq26} correspond to real excitations of each Mott lobe. Two of them are associated with stable quasiparticle and quasihole dispersions and the remaining two correspond to damped states. Thus, in order to proceed with the analysis, we must analyze \eref{eq26} in each Mott lobe with fixed integer particle density $n$ separately.

\subsection{Mott lobe $n=0$}

For the sake of simplicity, we consider first the case of $n=0$, i.e. for $\mu<-\Delta/2$. This case can be interpreted as a Mott lobe where the energy spectrum only contains the quasiparticle branch, and the energy gap depends on the potential barrier, i.e. it depends both on the tunneling energy and the disorder strength. Thus, we find that in this configuration, Eq.~\eref{eq26} admits two solutions. The first one corresponds to stable states
\begin{equation}\label{eq27}
\omega^+(\bi{k})=-\mu-\frac{\Delta}{2}\coth\Bigg(\frac{\Delta}{2J(\bi{k})}\Bigg),
\end{equation}
with diagonal components of the effective-mass tensor given according to \eref{eq20}
\begin{equation}\label{eq28}
m^*=\frac{8JD^2}{a^2\Delta^2}\sinh^2\Bigg(\frac{\Delta}{4JD}\Bigg).
\end{equation}
By taking the limit of $\Delta\rightarrow 0$ in \eref{eq27}, we recover the clean-case dispersion 
\begin{equation}\label{eq29}
\omega^+_0(\bi{k})=-\mu-J(\bi{k}).
\end{equation}
The second solution corresponds to damped states
\begin{equation}\label{eq30}
\omega_r(\bi{k})=-\mu-\frac{\Delta}{2}\tanh\Bigg(\frac{\Delta}{2J(\bi{k})}\Bigg),
\end{equation}
with a lifetime obtained from \eref{eq24}
\begin{equation}\label{eq31}
\tilde{\tau}(\bi{k})=\frac{4}{\pi\Delta}\cosh^2\Bigg(\frac{\Delta}{2J(\bi{k})}\Bigg).
\end{equation}
Using \eref{eq25}, we express the spectral function as 
\begin{equation}\label{eq32}
\eqalign{A(\bi{k},\omega)=&\frac{\Delta^2}{4J(\bi{k})^2}\mbox{csch}^2\Bigg(\frac{\Delta}{2J(\bi{k})}\Bigg)\delta(\omega-\omega^+(\bi{k}))
\\
& +
\frac{\frac{1}{\Delta}[\Theta (\omega+\mu+\frac{\Delta}{2})-\Theta (\omega+\mu-\frac{\Delta}{2})]}{\Bigg[1-\frac{J(\bi{k})}{\Delta}\mbox{log}\Big\vert\frac{\omega+\mu-\Delta/2}{\omega+\mu+\Delta/2}\Big\vert\Bigg]^2+\pi^2J(\bi{k})^2/\Delta^2}.}
\end{equation}
It can straightforwardly be checked that \eref{eq32} satisfies the general properties \eref{eq2}. Furthermore, the density of states follows from integrating \eref{eq32} over the first Brillouin zone
\begin{equation}\label{eq33}
\rho(\omega)=\int_{\rm BZ}\frac{d^Dk}{(2\pi)^D} A(\bi{k},\omega).
\end{equation}

\begin{figure}[!]
\centering
\includegraphics[width=\textwidth]{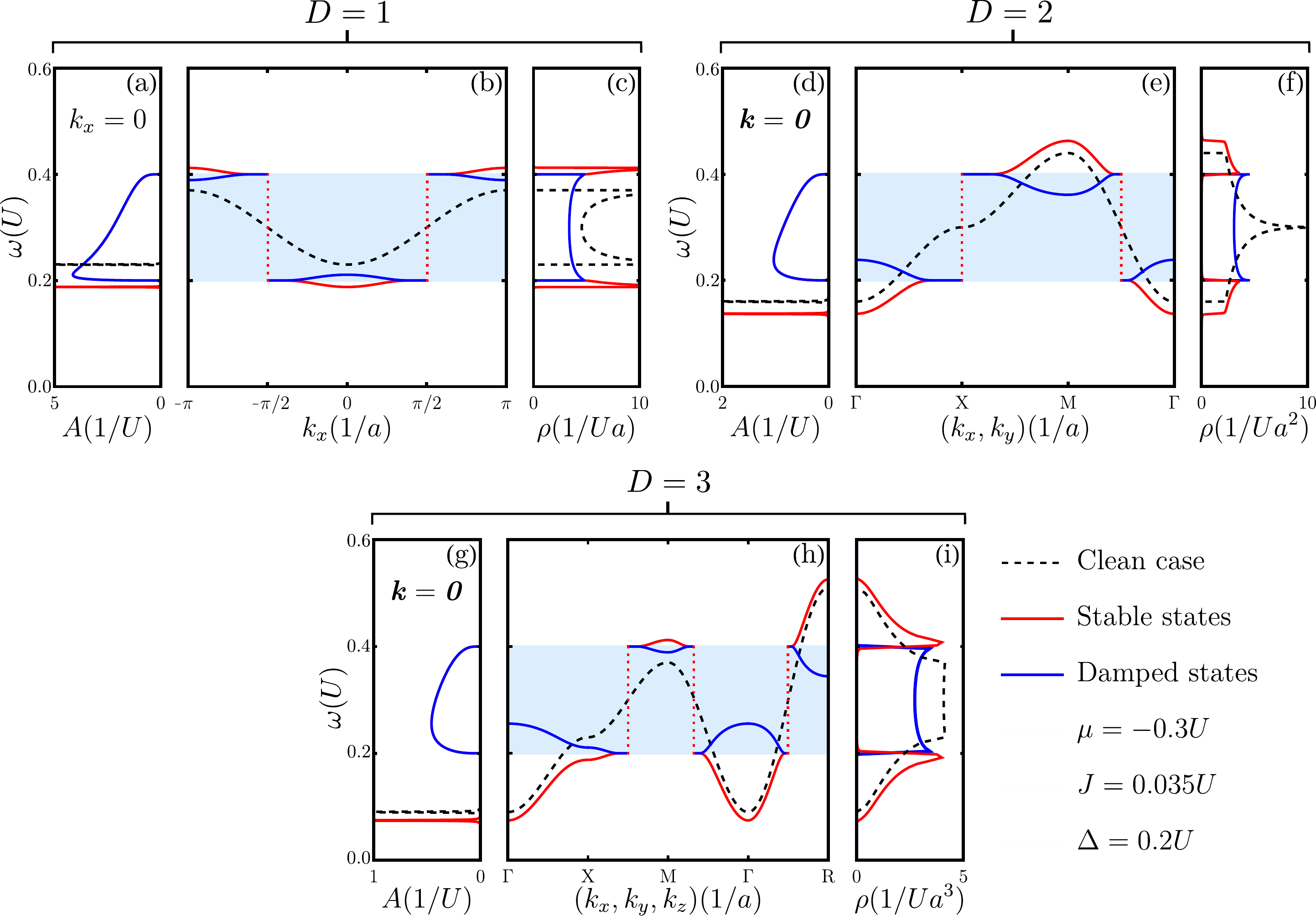}
\caption{\label{fig2}Band structure for the Mott lobe $n=0$. (a), (d) and (g) correspond to the spectral function \eref{eq32}. (b), (e) and (h) represent the dispersion relations \eref{eq27}, \eref{eq29} and \eref{eq30}. (c), (f) and (i) show the to the density of states \eref{eq33}. The shaded lightblue region and the dotted red lines in (b), (e) and (h) correspond to the band of damped states of width $\Delta$ and the jumps of the dispersion relations, respectively. Although interactions are not important in this case, we chose $U$ as the energy scale in order to specify the points in the $J\times\mu$ phase diagram.}
\end{figure}

In Fig.~\ref{fig2} we show a plot of the resulting band structure for $D=1$, $D=2$ and $D=3$. Although the results \eref{eq27}$-$\eref{eq33} for the Mott lobe $n=0$ do not depend on the interaction energy $U$ we use it as a measure for the energy scales in Fig.~\ref{fig2}. This allows to localize the equilibrium points in the phase diagram defined in the $J\times\mu$ plane. In Fig.~\ref{fig2}(a), (d) and (g) the special case of the spectral function for $\bi{k}=\textbf{\textit{0}}$ resembles the qualitative sketch depicted in Fig.~\ref{fig1}. We observe a sharp peak shifted due to disorder towards lower energies for the stable states (red) along with broad distribution for the damped states (blue). In Fig.~\ref{fig2}(b), (e) and (h) we plot \eref{eq27}, \eref{eq29}, and \eref{eq30}, where, for $D=2$, the critical points of the Brillouin zone are defined as $\Gamma=(0,0)$, X~$=(0,\pi)$, M~$=(\pi,\pi)$, while for $D=3$  such critical points are defined as $\Gamma=(0,0,0)$, X~$=(0,\pi,0)$, M~$=(\pi,\pi,0)$, and R~$=(\pi,\pi,\pi)$. There we observe the damped states inside the lightblue-shaded region of width $\Delta$. Note that the dispersions of both stable and damped states have jumps (dotted red lines). We can understand these jumps considering the random potential as a superposition of many spatial Fourier components. As the disorder is uncorrelated at different sites, no Fourier component with frequency larger than $\pi/a$ can exist.~Therefore, a matter wave excitation, which propagates, for instance, in the $k_x$ direction and interacts separately with each component of the disorder potential, would meet the condition for Bragg scattering exactly at $k_x=\pm\pi/2a$. Thus, we interpret these jumps as the scattering experienced by each wavevector component of the excitations interacting with the corresponding spatial frequency component of the disordered potential \cite{aspect2009anderson}. We point out that the group velocity described as the gradient of the dispersion relations, $\bi{v}_g=\nabla_{\bi{k}}\omega(\bi{k})$, vanishes at those points, which implies a localization of the wave packets. In accordance with the present results, previous studies \cite{roux2013dynamic} demonstrated the emergence of sub bands due to the disorder in the Bose-Hubbard model. In Fig.~\ref{fig2}(c), (f) and (i) we plot the density of states \eref{eq33}. In the $D=1$, Fig.~\ref{fig2}(c), we recognize that between the Van Hove singularities corresponding to the stable excitations (red), damped states (blue) emerge, thus increasing the band width to $\Delta\coth(\Delta/4J)$. In $D=2$ the strong peak that appears in the clean-case band center, gets divided into two peaks exactly at the frequencies where the dispersion of the stable states meets the dispersion of the damped states, namely at $\omega=-\mu\pm\Delta/2$. An analogous situation occurs in $D=3$.

\begin{figure}[!]
\centering
\includegraphics[width=.75\textwidth]{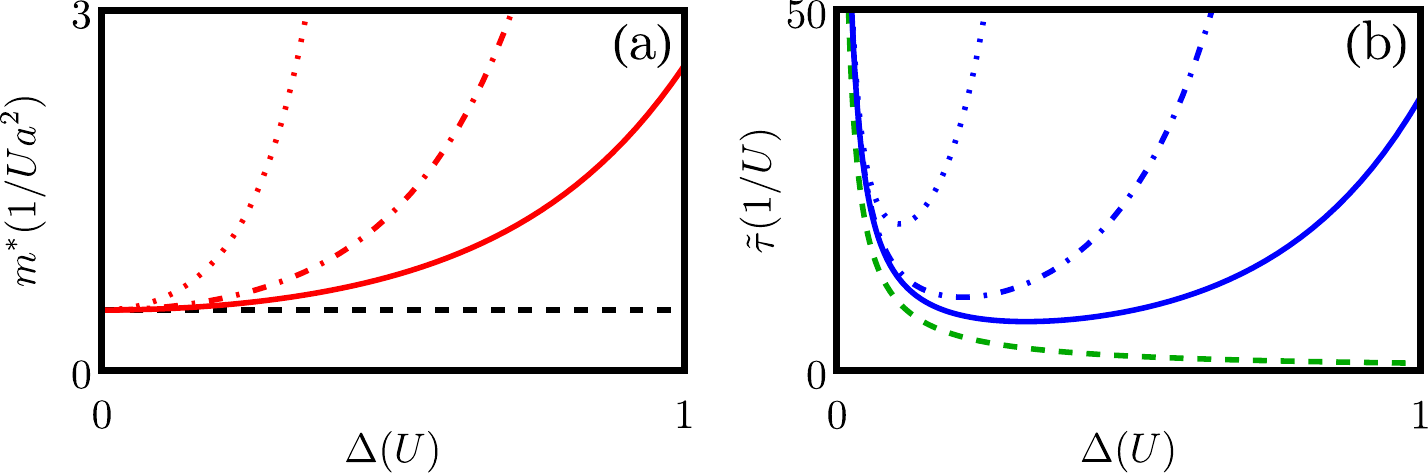}
\caption{\label{fig3} (a) Effective mass \eref{eq27}. (b) Lifetime at $\bi{k}=\textbf{\textit{0}}$ \eref{eq30}. In both plots the dotted, dot-dashed and continuous red and blue lines correspond to $D=1$, $D=2$ and $D=3$, respectively. The dashed black line in (a) and the dashed-green line in (b) correspond to the clean-case effective mass and the time associates in the inverse of energy scale of disorder $1/\Delta$, respectively. We use here the same parameters as in Fig~\ref{fig2}.}
\end{figure}

In order to understand what happens for increasing disorder, we plot the effective mass of the stable states and the lifetime of the damped states in Fig.~\ref{fig3}. In Fig.~\ref{fig3}(a) we find that the effect of disorder is to increase the effective mass of the stable states, thus indicating that their dispersion is becoming more flat. In  Fig.~\ref{fig3}(b), we plot the lifetime of the damped states. We deduce from such a plot that the lifetime (blue) increases for large $\Delta$, which corresponds to the resonance becoming sharply peaked. Thus, these states become more stable when the disorder strength $\Delta$ is of the order of the interaction energy $U$. 

We make the connection to the quantum phase transitions by analyzing the excitations gap. For damped states the gap is given by the lower bound of the broad distribution in \eref{eq32}, namely $E^r_0=-\mu-\Delta/2$, while the gap for stable states is obtained by expanding \eref{eq27} near $\bi{k}=\textbf{\textit{0}}$, yielding $E^+_0=-\mu-\frac{\Delta}{2}\coth(\Delta/4JD)$. For sufficiently small tunneling, we get that $E^r_0-E^+_0\sim \Delta{\rm e}^{-\Delta/2JD}$, which is always positive. Thus, increasing $\Delta$ has the effect that the stable states gap closes before the gap for damped states. However, in this limit, the dispersive  nature of these excitations disappears indicating a broad distribution in momentum, which is characteristic of localized states. For sufficiently strong disorder, the damped sates occupy the whole band. This corresponds to a transition from Mott to Bose glass. Therefore, we come to the fundamental conclusion that the damped states correspond to single-particle excitations of the Bose-glass state. In the presence of disorder no direct Mott-superfluid transition is possible \cite{fisher1989boson,rapsch1999density,pollet2009absence,gurarie2009phase}. In order to get information on the complete quantum phase diagram, one would have to consider higher number of scattering processes \cite{souza2021green}. Quantum and thermal fluctuations could also be included using an effective-action approach \cite{bradlyn2009effective}. It has been proposed that one could define fluctuations in the disorder average of the mean particle density as an order parameter to identify the Bose-glass phase, in analogy to the Edwards-Anderson order parameter in the spin glass theory \cite{graham2009order,morrison2008physical}. The incorporation of such an order parameter could lead to precise results inside the Bose-glass phase.

Next, we analyze the case of $n=1$, where interactions become important.

\subsection{Mott lobe $n=1$}

In the case of $n=1$, assuming that the arguments of the absolute values are real, \eref{eq26} simplifies to  
\begin{equation}\label{eq34}
\fl
\Big(\omega+\mu-U-\frac{\Delta}{2}\Big)^2\Big(\omega+\mu+\frac{\Delta}{2}\Big)
+\xi\mathrm{e}^{\frac{\Delta}{J(\bi{k})}}\Big(\omega+\mu+\frac{\Delta}{2}\Big)^2\Big(\omega+\mu-\frac{\Delta}{2}\Big)
=0,
\end{equation}
where $\xi=\pm 1$. Due to  the different conditions imposed by a uniform disorder distribution on the frequencies that satisfy \eref{eq34}, we distinguish the intervals
\begin{equation}\label{eq35}
\eqalign{
-\mu-\Delta/2\quad & <\quad \omega\quad<\quad-\mu+\Delta/2,
\\
U-\mu-\Delta/2\quad &<\quad \omega\quad <\quad U-\mu+\Delta/2.
}
\end{equation}
Real solutions of \eref{eq34} inside these intervals correspond to damped states. On the other hand, frequencies that satisfy \eref{eq34} outside the intervals \eref{eq35} correspond to stable states. 

We note that \eref{eq34} can be rewritten as a cubic equation of the form 
\begin{equation}\label{eq36}
\omega^3+B_2\omega^2+B_1\omega+B_0=0,
\end{equation}
which admits three solutions for each value of $\xi$. 
\begin{figure}[!]
\centering
\includegraphics[width=.85\textwidth]{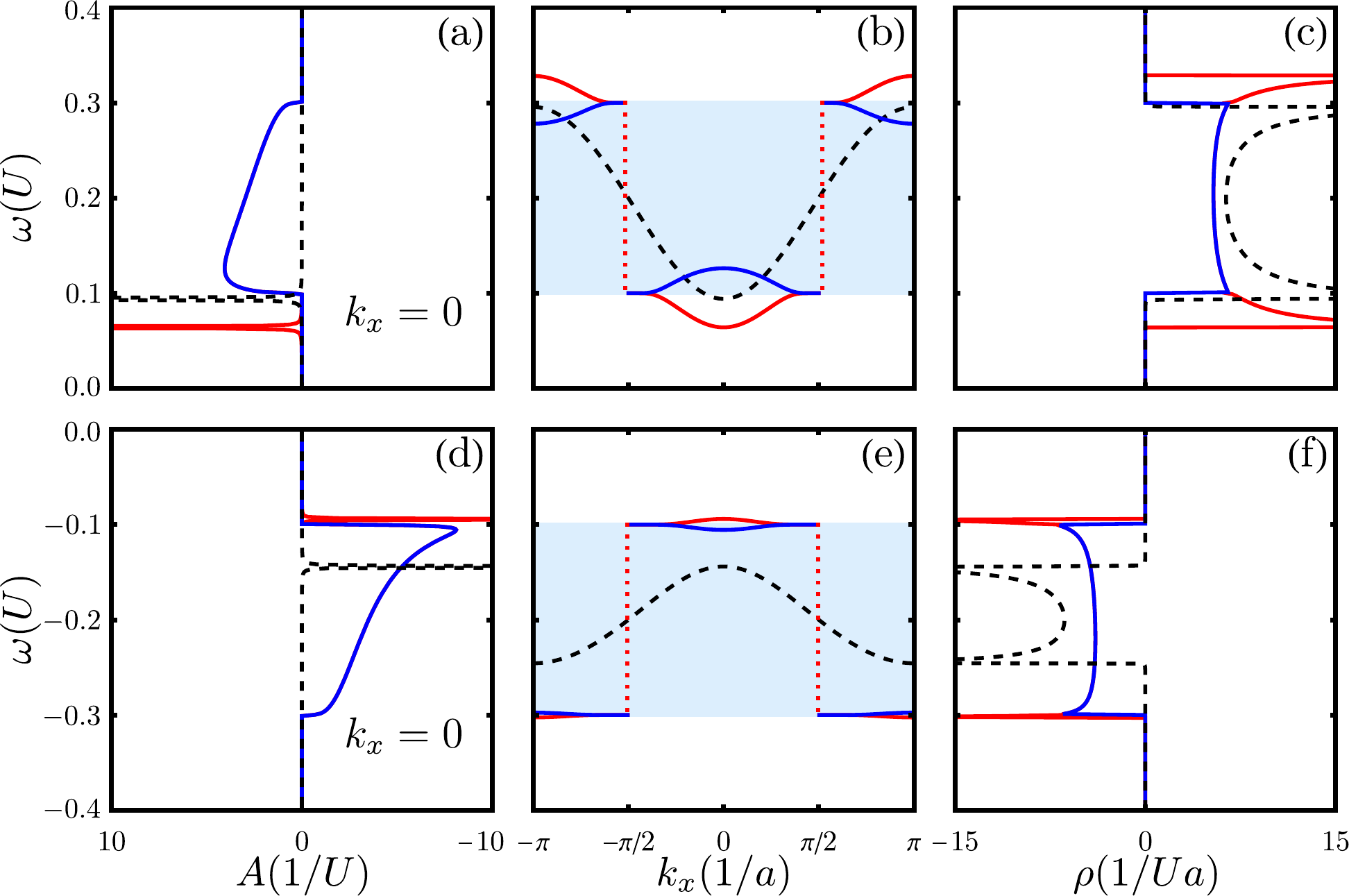}
\caption{\label{fig4}Band structure for the Mott lobe $n=1$ in the case of $D=1$. The upper row depicts to the quasiparticle branch with $\mu=0.8U$ and the lower row corresponds to the quasihole branch $\mu=0.2U$. (a) and (d) represent to the spectral function \eref{eq25}. (b) and (e) follows from the dispersion relations \eref{eq36}. (c) and (f) results from the to the density of states \eref{eq33}. The shaded lightblue region and the dotted red lines in (b) and (e) correspond to the band of damped states of width $\Delta$ and the jumps of the dispersion relations, respectively. In all plots we have chosen $J=0.025U$ and $\Delta=0.2U$.}
\end{figure}
Using the reduced form for this cubic equation and applying Cardano's formula, we get the solutions in the form
\begin{equation}\label{eq37}
\omega_l=-\frac{B_2}{3}+\alpha^{l}C-\frac{B_1-\frac{{B_2}^2}{3}}{\alpha^{l}C},
\end{equation}
where $\alpha=\frac{-1+\mathrm{i}\sqrt{3}}{2}$ is the primitive cubic root of unity and we have introduced the abbreviations
\begin{equation}\label{eq38}
\fl
C=\Bigg[\frac{1}{2}\Bigg(\frac{B_1B_2}{3}-B_0-\frac{2B_2^3}{27}\Bigg)
+\sqrt{\frac{1}{4}\Bigg(B_0-\frac{B_1B_2}{3}+\frac{2B_2^3}{27}\Bigg)^2+\frac{1}{27}\Bigg(B_1-\frac{B_2^2}{3}\Bigg)^3}\Bigg]^{\frac{1}{3}},
\end{equation}
\begin{equation}\label{eq39}
\fl
B_2=3\mu-2U+\frac{\Delta}{2}\Bigg(\frac{-1+\xi{\rm e}^{\frac{\Delta}{J(\bi{k})}}}{1+\xi{\rm e}^{\frac{\Delta}{J(\bi{k})}}}\Bigg),
\end{equation}
\begin{equation}\label{eq40}
\fl
B_1=U^2-\frac{\Delta^2}{4}-4U\mu+3\mu^2+\Delta\mu \Bigg(\frac{-1+\xi{\rm e}^{\frac{\Delta}{J(\bi{k})}}}{1+\xi{\rm e}^{\frac{\Delta}{J(\bi{k})}}}\Bigg),
\end{equation}
\begin{equation}\label{eq41}
\fl
B_0=\mu U^2-2U\mu^2+\mu^3+\frac{U\Delta^2}{2}-\frac{\mu\Delta^2}{4}
+\Bigg(\frac{\Delta\mu^2}{2}-\frac{\Delta U^2}{2}-\frac{\Delta^3}{8}\Bigg)\Bigg(\frac{-1+\xi{\rm e}^{\frac{\Delta}{J(\bi{k})}}}{1+\xi{\rm e}^{\frac{\Delta}{J(\bi{k})}}}\Bigg).
\end{equation}
Note that the on-site interaction energy $U$ appears explicitly for $n=1$ according to Eqs.~\eref{eq39}$-$\eref{eq41} in contrast to the $n=0$ case treated in Sec.~4.1. The dispersion relation of each excitation is obtained from \eref{eq37} considering the different exponents of the primitive cubic root defined as  $l\in \{0,1,2\}$. For the case of $\xi=-1$ all three solutions are real. However, for $\xi=1$ only the solution for $l=1$ is real. Therefore, these are the solutions of \eref{eq36} which correspond to excitations of the $n=1$ Mott lobe.   

In order to illustrate the results, we plot in Fig.~\ref{fig4} the $D=1$ band structure for the $n=1$ Mott lobe. We observe that such a plot resembles qualitatively the $n=0$ case in Fig.~\ref{fig2}, i.e., the gap for stable states decreases and damped states emerge in the middle of the band. Additionally, as a result of the scattering with the random potential, the dispersions have jumps exactly at $k_x=\pi/2a$ analogously to the previous $n=0$ case. However, instead of having only the quasiparticle branch, in this case the band structure shows for negative energies a quasihole branch as well.
\begin{figure}[!]
\centering
\includegraphics[width=.75\textwidth]{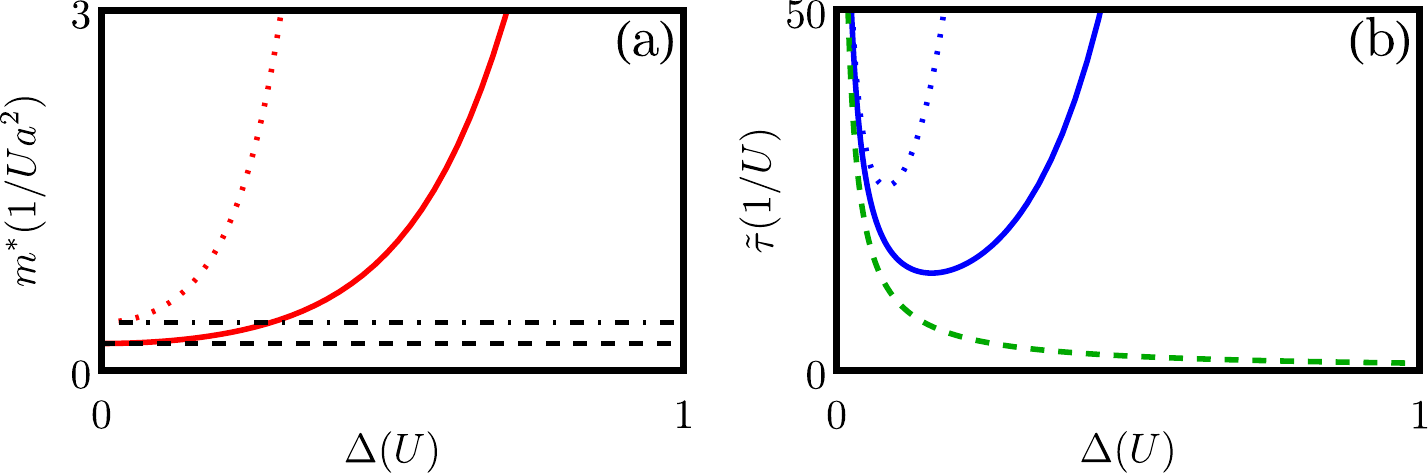}
\caption{\label{fig5} (a) Effective mass \eref{eq20}. (b) Lifetime at $\bi{k}=\textbf{\textit{0}}$ \eref{eq24}. In both plots the dotted and continuous red and blue lines correspond to the quasihole and quasiparticle branches, respectively. The dashed and dot-dashed black lines in (a) correspond respectively clean-case quasiparticle and quasihole effective masses. The dashed green line in (b) corresponds to the time scale associated to the inverse of the energy scale of disorder, $1/\Delta$. In both plots we chose $J=0.025U$.}
\end{figure} 

Applying equations \eref{eq20} and \eref{eq24}, we plot in Fig.~\ref{fig5} the effective mass and the lifetime corresponding to the stable and damped states in the quasiparticle and quasihole branches, respectively. We observe that, analogously to the previous case of $n=0$ in Fig.~\ref{fig3}, the effective mass of the stable states as well as the lifetime of the damped states increase in the strong disorder limit. We read off from Fig.~\ref{fig5} that both the effective mass of the stable states and the lifetime of the damped states in the quasihole branch (doted red line in (a) and dotted blue line in (b)) depend more sensibly upon the disorder strength.

It is important to note what happens when the gap for creating these excitations closes. In our present case, the excitation spectrum comprises both the quasiparticle and quasihole branches. During a generic phase transition, the gap of one of these branches will close while the gap for the other remains open. Restricting the analysis to the quasiparticle branch of the spectrum, the gap for stable excitations can be computed by expanding \eref{eq37} near $\bi{k}=\textbf{\textit{0}}$ for $l=1$ and $\xi=-1$. In the asymptotic limit of vanishing tunneling energy, such a gap can be expressed as $E_1^+\sim U-\mu-\Delta(1/2+{\rm e}^{-\Delta/2JD})$. The gap for damped states, however, can be computed from the left-hand side of the second line of \eref{eq35}, which yields $E_1^r=U-\mu-\Delta/2$. Analogously to the case analyzed in Section 4.1, the difference between the two energy gaps reads $E_1^r-E_1^+\sim\Delta{\rm e}^{-\Delta/2JD}$, which is again always positive. By applying the same reasoning as previously discussed, we observe that increasing $\Delta$ causes the gap for stable states to close before the gap for damped states. However, in this scenario, as depicted in Fig.~\ref{fig5}, the effective mass of stable excitations increases, and their dispersive nature vanishes, indicating a broad distribution in momentum, which is a characteristic of localized states. As the disorder strength becomes sufficiently high, the damped states fill the entire band, leading to a transition from Mott to Bose glass. This confirms our fundamental conclusion that the damped states correspond to single-particle excitations of the Bose-glass state. In the clean case, it was shown that in the first Mott lobe, when the quasiparticle gap closes during a generic phase transition, it transforms continuously into a Goldstone mode of the superfluid phase, while the quasihole gap, which remains open, transforms continuously into a gapped amplitude mode \cite{grass2011excitation}. We, therefore, expect that this scenario will also hold when disorder is present, such that the stable and damped excitations will transform continuously into excitations of the superfluid phase. However, further analysis is required to draw definitive conclusions regarding the effects of disorder on these superfluid excitations. Our next area of focus is to examine the spatio-temporal propagation of stable and damped excitations.

\section{Spatio-temporal profile of the Green's function}

Thus far, we have focused on the properties of the available states for an excitation in Fourier space. We now turn our attention to the implications of the findings demonstrated on the previous section in real space and time. To this end, we investigate the full Green's function within our approximation.

First, we obtain the long-wavelength behavior of the Green's function in space by integrating \eref{eq15} over the first Brillouin zone in the limit of large space separations, which yields
\begin{equation}\label{eq41}
\langle\mathcal{G}_{ij}(\omega)\rangle\sim\frac{1}{|\bi{r}_i-\bi{r}_j|^{\frac{D-1}{2}}}\exp\Bigg(-\frac{|\bi{r}_i-\bi{r}_j|}{\ell}\Bigg)\quad\textrm{as}\quad|\bi{r}_i-\bi{r}_j|\rightarrow\infty,
\end{equation} 
where the length scale associated with the exponential decay of $\langle\mathcal{G}_{ij}(\omega)\rangle$ turns out to be
\begin{equation}\label{eq42}
\ell=a\frac{\sqrt{J|\langle {\cal G}(\textbf{\textit{0}},\omega)\rangle|}}{|\sin[\arg(\langle {\cal G}(\textbf{\textit{0}},\omega)\rangle)/2]|}.
\end{equation}
This corresponds to the mean free path of excitations between each scattering event \cite{economou2006green}. Note that only the states at $\bi{k}=\textbf{\textit{0}}$ contribute to the asymptotic behavior of $\langle\mathcal{G}_{ij}(\omega)\rangle$ for large separations. The exponential decay of the single-particle Green's function is a general consequence of the mass gap present in the Mott insulating phase \cite{fradkin2021quantum}. However, the mean free path diverges as $\mbox{Im}\langle g_i(\omega)\rangle\rightarrow 0$, so the amplitude of propagation for stable states decays algebraically as $|\bi{r}_i-\bi{r}_j|^{-(D-1)/2}$. For finite $\mbox{Im}\langle g_i(\omega)\rangle$, i.e., in the energy range of damped states, there is an additional exponential decay with the characteristic length $\ell$ defined in \eref{eq42}. We remark that, within our approach, the asymptotic decay of $\langle\mathcal{G}_{ij}(\omega)\rangle$ given in \eref{eq41} is generally valid for all Mott lobes and any form of bounded disorder distribution $p(\epsilon_i)$. 

\begin{figure}[!]
\centering
\includegraphics[width=1\textwidth]{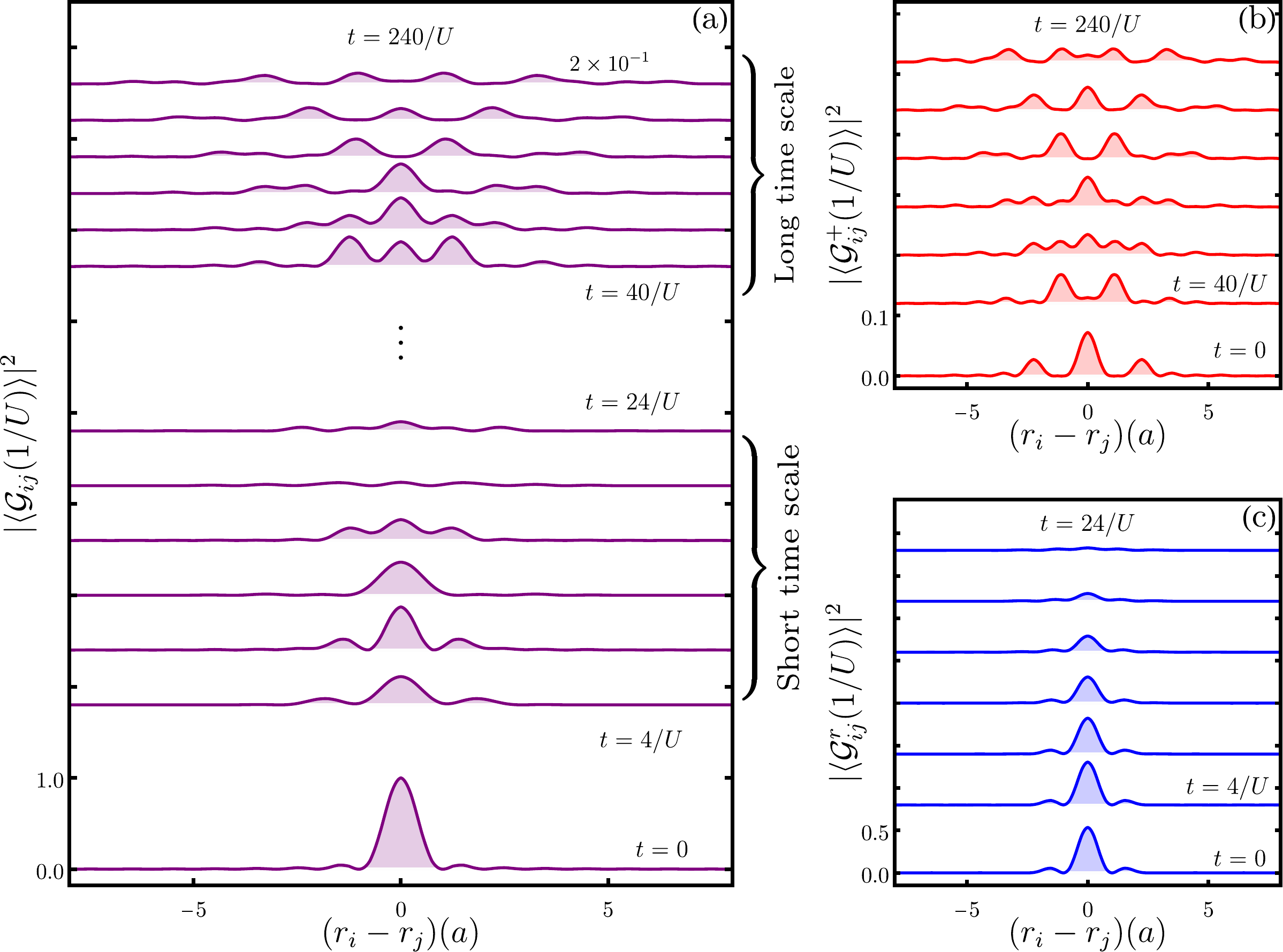}
\caption{\label{fig6} Time evolution of the absolute square value of the Green's function for $D=1$ and $n=0$. (a) correspond to the total Green's function $|\langle\mathcal{G}_{ij}(t)\rangle|^2$, where we have used feature scaling better illustrate the behavior of the amplitude at long time scales. (b) Contribution $|\langle\mathcal{G}_{ij}^+(t)\rangle|^2$ of the stable states. (c) Contribution $|\langle\mathcal{G}_{ij}^r(t)\rangle|^2$ of the damped states. In all plots we chose $J=0.035U$, $\Delta=0.2U$ and $\mu=-0.3U$. }
\end{figure}

In order to get the complete picture, we write the retarded Green's function in terms of the spectral function by using the following representation \cite{fetter2012quantum}
\begin{equation}
\langle {\cal G}(\bi{k},\omega)\rangle=\int^\infty_{-\infty}d\omega^\prime\frac{A(\bi{k},\omega)}{\omega-\omega^\prime+i0^+}.
\end{equation}
Hence, the spatio-temporal profile can be written as
\begin{equation}\label{eq43}
\langle\mathcal{G}_{ij}(t)\rangle=\mathrm{i}\Theta(t)\Bigg(\frac{a}{2\pi}\Bigg)^D\int_{\rm BZ}d^Dk\int^{\infty}_{-\infty}d\omega 
A(\bi{k},\omega)\mathrm{e}^{-i\omega t+i\bi{k}\cdot(\bi{r}_i-\bi{r}_j)}.
\end{equation}
For $t=0$, the integration over the frequency domain becomes the sum rule \eref{eq2}. In this case, solving the integral in $k$ yields
\begin{equation}\label{eq44}
\langle\mathcal{G}_{ij}(0)\rangle=\prod_{q=1}^D\frac{\sin(\pi r_{ij}^{(q)}/a)}{\pi r_{ij}^{(q)}/a},
\end{equation}
where $r_{ij}^{(q)}$ is the $q$-ht component of the vector $\bi{r}_i-\bi{r}_j$. Note that this result is the same for the clean and the disordered cases.
Furthermore, using \eref{eq32} for the case of $n=0$ we distinguish for $t>0$ the contributions
\begin{equation}\label{eq45}
\fl
\langle\mathcal{G}_{ij}^+(t)\rangle=\mathrm{i}\Theta(t)\Bigg(\frac{a}{2\pi}\Bigg)^D\int_{\rm BZ}d^Dk\frac{\Delta^2}{4J(\bi{k})^2}\mbox{csch}^2\Bigg(\frac{\Delta}{2J(\bi{k})}\Bigg)\e^{-i\omega^+(\bi{k})t},
\end{equation}
\begin{equation}\label{eq46}
\fl
\langle\mathcal{G}_{ij}^r(t)\rangle=\mathrm{i}\Theta(t)\Bigg(\frac{a}{2\pi}\Bigg)^D\int_{\rm BZ}d^Dk\int^{\infty}_{-\infty}d\omega\frac{\Delta^2 p(\omega+\mu)\mathrm{e}^{-i\omega t+i\bi{k}\cdot(\bi{r}_i-\bi{r}_j)}}{\Delta^2\Bigg[1-\frac{J(\bi{k})}{\Delta}\mbox{log}\Big\vert\frac{\omega+\mu-\Delta/2}{\omega+\mu+\Delta/2}\Big\vert\Bigg]^2+\pi^2J(\bi{k})^2}
\end{equation}
corresponding to the stable and damped states, respectively. 

The absolute squared value of these quantities is plotted in Fig.~\ref{fig6}. In Fig.~\ref{fig6}(a) we observe two distinct regimes for the time evolution of $|\langle\mathcal{G}_{ij}(t)\rangle|^2$. At short time scales its amplitude is localized around $r_i-r_j=0$. However, at time $t=16/U$ the amplitude starts to become extended. For long time scales only the extended states contribute to $|\langle\mathcal{G}_{ij}(t)\rangle|^2$. In Fig.~\ref{fig6}(b) we read off that the contribution $\langle\mathcal{G}_{ij}^+(t)\rangle$ of the stable states spreads through space as time increases. Therefore, in the long-time limit, there exists a finite probability of finding the excitation on a site arbitrarily distant from the site where it was created. We conclude from \eref{eq41} that the algebraic decay is not enough to localize theses excitations in space. In Fig.~\ref{fig6}(c) we notice no diffusion of the contribution $\langle\mathcal{G}_{ij}^r(t)\rangle$ together with a rapid decay in time. Thus, the additional exponential decay contributes to the localization of the damped states. Therefore, the behavior of the full Green's function in space is dominated by the damped states at short time scales, while for long time scales it is dominated by the stable states.

\section{Summary and conclusions}

In conclusion, we investigated the effect of disorder on the low-energy excitations of the Bose-Hubbard model in the strongly interacting limit at zero temperature applying a perturbative field-theoretical approach to obtain a resummed expression for the spectral function. By analyzing the peaks of the spectral function we demonstrated that two different kinds of excitation states are present, namely stable states which are extended in space with algebraically decaying amplitude and damped states which are localized with exponentially decaying amplitude. By considering the limit of strong disorder, where the damped states dominate the spectrum with lifetime increased by the disorder, we argued that they correspond to low-energy single-particle excitations of the Bose-glass phase. Our results inside each Mott lobe for small values of the tunneling energy are general, and therefore independent of the exact form of the bounded disorder distributions. Furthermore, by analyzing the case of uniform distribution we showed that disorder increases the effective mass of the stable states. We point out that the spectral function can experimentally be determined, for instance by Bragg spectroscopy \cite{clement2010bragg,fabbri2012quasiparticle} or by the radio frequency transfer method \cite{volchkov2018measurement}. Moreover, it has been recently proposed that the spectral function could be used to probe key aspects of the excitation spectrum in the disordered case in quench spectroscopy experiments \cite{villa2021quench,villa2021finding}. Further insight into the transition to the superfluid phase could be obtained by including loop corrections to the resummation method developed here. Future research is required to determine the role of the damped states on the multiple matter wave interference pattern which can be measured in time-of-flight experiments. It was demonstrated that the effect of spatially random fields in bosonic ultracold quantum gases can be mimicked by a time-alternating external potential resulting in a Bose-glass like nonequilibrium granular condensate \cite{yukalov2009bose}. Since it has been reported that a free expanding out-of-equilibrium condensate resembles a propagating optical speckle \cite{tavares2017matter}, exploring the connection between the effects of spatially random and time-varying potentials could be an interesting direction for future work.   

\ack
We thank A. Balaž, M. Bonkhoff, H. Kroha and F. B. Ramos for
comments that greatly improved the present work. We acknowledge the support
of the Coordenação de Aperfeiçoamento de Pessoal de Nível Superior (CAPES) and
the Deutscher Akademischer Austauschdienst (DAAD) under the bi-national joint
program CAPES-DAAD PROBRAL Grant number 88887.627948/2021-00.  R. S. S.
acknowledges the funding support of CAPES under Programa de Demanda Social
(Social Demand Program) grant number 88882.426685/2019-01. A. P. acknowledges
financial support by the Deutsche Forschungsgemeinschaft (DFG) via the Collaborative
Research Center SFB/TR185 (Project No. 277625399).

\section*{References}

\bibliographystyle{unsrt}
\bibliography{references}

\end{document}